\documentclass[aps,prl,reprint,superscriptaddress]{revtex4-1}

\usepackage{amssymb,amsmath}
\usepackage{booktabs}
\AtBeginDocument{
	\heavyrulewidth=.08em
	\lightrulewidth=.05em
	\cmidrulewidth=.03em
	\belowrulesep=.65ex
	\belowbottomsep=0pt
	\aboverulesep=.4ex
	\abovetopsep=0pt
	\cmidrulesep=\doublerulesep
	\cmidrulekern=.5em
	\defaultaddspace=.5em
}
\def\br{{\bf r}}
\def\sss{\scriptscriptstyle\rm}

\def\VWa{vW}

\def\half{\frac{1}{2}}
\def\intdhrs{\int d^3r}

\def\hatT{\hat{T}}
\def\hatV{\hat{V}}

\def\nad{^{\sss{nad}}}
\def\VW{^{\sss{vW}}}

\def\f{_{\sss f }}
\def\p{_{\sss p}}
\def\s{_{\sss s}}
\def\H{_{\sss H}}
\def\xc{_{\sss{xc}}}
\def\ee{_{\sss{ee}}}
\def\ext{_{\sss{ext}}}

\begin{document}

\title{Virial Relations in Density Embedding}

\author{Kaili Jiang}
\affiliation{Department of Chemistry, 73 Warren St., Rutgers University, Newark, NJ 07102, USA}

\author{Mart\'{\i}n A. Mosquera}
\affiliation{Department of Chemistry, Northwestern University, 2145 Sheridan Road, Evanston, Illinois 60208, USA}

\author{Yan Oueis}
\affiliation{Department of Chemistry, Purdue University, 560 Oval Dr., West Lafayette IN 47907, USA}

\author{Adam Wasserman}
\email[Corresponding Author: ]{awasser@purdue.edu}
\affiliation{Department of Chemistry, Purdue University, 560 Oval Dr., West Lafayette IN 47907, USA}
\affiliation{Department of Physics and Astronomy, Purdue University, 525 Northwestern Ave., West Lafayette, IN 47907, USA}

\begin{abstract}
The accuracy of charge-transfer excitation energies,
solvatochromic shifts and other environmental effects calculated via various density embedding techniques depend critically on the approximations employed for the non-additive non-interacting kinetic energy functional, $T_{\sss s}^{\rm nad}[n]$. Approximating this functional remains an important challenge in electronic structure theory. To assist
in the development and testing of approximations for $T_{\sss s}^{\rm nad}[n]$,
 we derive two virial relations
for fragments in molecules. These establish separate connections between
the non-additive kinetic energies of the non-interacting and interacting systems of
electrons, and quantities such as the electron-nuclear attraction forces, the partition (or
embedding) energy and potential, and the Kohn-Sham potentials of the system and its
parts. We numerically verify both relations on diatomic molecules. 

\end{abstract}%

\maketitle

\section{Introduction}

Solving the electronic-structure problem for ever larger, complex molecular systems, has prompted the development of quantum embedding methods \cite{SC16}.  Among these, density-embedding techniques \cite{NW14} offer the most direct way to fragment a composite system (molecule\cite{GORW19}, cluster\cite{HLPC14}, material\cite{GCP14}) into its constituent parts while making direct use of Density Functional Theory (DFT) \cite{HK64,KS65} and its popular approximations. Frozen-density embedding theory (FDE) \cite{WW93}, subsystem-DFT (S-DFT) \cite{KSGP15}, and Partition-DFT (P-DFT) \cite{CW07,EBCW10} belong to this category of embedding methods. To be competitive with KS-DFT, one feature that these three methods have in common is their reliance on approximations for the non-additive non-interacting kinetic energy $T\nad_{\sss s}$ (NAKE), a quantity that can be thought of as a functional of the total ground-state density or, alternatively, of the set of fragment ground-state densities. Other density-embedding methods circumvent this need altogether at the expense of an initial supra-molecular calculation \cite{MSGMI12}. The development of accurate approximations of the full $T_{\sss s}[n]$ for orbital-free DFT is a notoriously difficult problem \cite{WW13}, explaining why most DFT calculations today still rely on the Kohn-Sham (KS) \cite{KS65} or generalized-KS \cite{KK08} schemes.  However, approximating the NAKE is a different challenge than approximating the full $T_{\sss s}[n]$. Cancellation of errors can sometimes lead to acceptable NAKEs \cite{GCP16} but not much is known about such errors or how to control them. Deriving exact conditions for the NAKE would be helpful to guide the construction of improved approximations for it \cite{LKW08}. We derive here two virial relations that may be useful toward that goal.                                                                                                                                                                                                                                                                                                                                                                                                                                                                                                                                                                                                                                                

The quantum virial theorem provides relationships between the kinetic energy and the potential energy of electronic systems.
In Kohn-Sham DFT \cite{KS65}, virial relations have been proven \cite{LP85,GS89,RAGC09} that establish the connections between the kinetic and potential energies of both, the real system of interacting electrons and the auxiliary system of non-interacting electrons. 
Establishing analogous virial relations in embedding methods is challenging when the fragment densities are not $v$-representable \cite{SB75,B86}, as discussed in ref.\cite{LP86} and in ref.\cite{HH75} in the context of the early embedding method of ref.\cite{GK72}.  However, the fragment densities of P-DFT {\em are} physical ground-state $v$-representable densities for which virial relations apply just as they would for any physical system in isolation. Furthermore, due to the globality of the partition potential in P-DFT \cite{CW07,EBCW10,NW14}, terms can be grouped together leading to particularly simple virial expressions, as we show here.

We make use of the following notation: Using the index ``$\alpha$" to label the fragments, the kinetic energy of fragment $\alpha$ is $K_{\alpha}[n_{\alpha}]$. The non-additive kinetic energy is defined as:
\begin{equation}\label{eqn:Knad_def}
K^{\rm nad}[\{n_{\alpha}\}] \equiv K[n]-\sum_{\alpha}K_{\alpha}[n_{\alpha}]~~,
\end{equation}
where $K[n]$ is the total kinetic energy for density $n(\br)$. Equation \ref{eqn:Knad_def} is the most direct method to calculate $K^{\rm nad}$. We will be contrasting Equation \ref{eqn:Knad_def} later on with a virial expression, Equation \ref{eqn:virial_pdft}. Similarly, the NAKE is defined by 
\begin{equation}
T\nad_{\sss s}[\{n_{\alpha}\}]\equiv T_{\sss s}[n]-\sum_\alpha T_{\sss s}[n_\alpha]~~,
\label{e:Ts_I}
\end{equation}

With the virial theorem, we can derive exact relations between $T_s^{\rm nad}[n]$ and the densities and potentials that can be obtained through P-DFT calculations. These relations can be used as exact constraints in constructing approximations to $T_s^{\rm nad}[n]$.

This manuscript is divided as follows: In Section \ref{sec:2} we derive two virial relations for P-DFT. The
first, Equation \ref{eqn:virial_pdft}, expresses an exact relationship between the non-additive kinetic energy of the
systems of interacting electrons $K^{\rm nad}[n]$, electrostatic electron-nuclear attraction energies, and the
partition energy $E_p[n]$ and corresponding potential $v_p(\br)$ \cite{EBCW10}. The second virial expression, Equation \ref{eqn:virial_pdft_alt}, relates $T_s^{\rm nad}[n]$ with the set of Kohn-Sham potentials for the fragments and for the whole system. Finally, in Section \ref{sec:3}, we provide numerical
verification of the derived relations for several homonuclear diatomic molecules, and discuss a few implications.

\section{Virial Relations}
\label{sec:2}
We now derive two virial relations for fragments in molecules. For a many-electron system of
ground state $|\Psi\rangle$ and density $n(\br)=\langle \Psi|\hat{n}(\br)|\Psi\rangle$ governed by the hamiltonian 
\begin{equation}
\hat{H}=\hat{T}+\hat{V}\ee+\int d^3r \hat{n}(\br)v(\br)~~,
\end{equation}
where $v(\br)$ is the `external' potential due to the nuclei, the virial theorem can be expressed as \cite{LP85}:
\begin{equation}\label{eqn:k}
2K[n]+V\ee[n]=\intdhrs n(\br)\br\cdot\nabla v(\br)~~,
\end{equation}
where $K[n]=\langle{\Psi[n]}|\hatT|{\Psi[n]}\rangle$ and $V\ee=\langle{\Psi[n]}|\hat{V}\ee|{\Psi[n]}\rangle$ are the total kinetic and electron repulsion energies. 
Similarly, using standard DFT notation for the KS system of non-interacting electrons with kinetic energy $T_s$,  
\begin{equation}\label{eqn:virial_KS}
2T\s[n]=\intdhrs n(\br)\br\cdot\nabla v\s[n](\br)~~,
\end{equation}
where the total KS potential $v\s[n](\br)$ is given by the sum of the external $v(\br)$, Hartree $v\H[n](\br)$, and exchange-correlation $v\xc[n](\br)$ potentials. Equation \ref{eqn:virial_KS}
is applicable not only to the exact XC functional, but also to approximate XC functionals at self-consistency.

Although P-DFT makes use of a grand-canonical ensemble formalism to describe fragments with fractional numbers of electrons \cite{NW14,NJW17}, we restrict the present analysis for simplicity to cases where the fragments, labeled by index $\alpha$, have integer numbers of electrons $N_\alpha$ (the one exception in the examples that follow is H$_2^+$, where each atomic fragment is assigned a charge of 0.5). The total number of electrons in the molecule, $N$, is given by the sum of the $N_\alpha$, and all single-particle operators are similarly additive. In particular, the external potential $v(\br)=\sum_\alpha v_\alpha(\br)$, kinetic operator $\hat{T}=\sum_\alpha \hat{T}_\alpha$ and density operator $\hat{n}(\br)=\sum_\alpha \hat{n}_\alpha (\br)$ are all additive. However, $\hat{V}\ee\neq\sum_\alpha\hatV_{\sss{ee},\alpha}$ as all electrons interact with one another.

Partition Theory establishes \cite{CW07,EBCW10} that there is only one embedding potential $v_p(\br)$ such that the many-electron Schr\"{o}dinger equations 
\begin{equation}
\label{eqn:Fragment-Schroedinger}
\left[\hat{H}_\alpha+\intdhrs v\p(\br)\hat{n}_\alpha(\br)\right]|{\psi_{\alpha}}\rangle=E_\alpha|{\psi_{\alpha}}\rangle~~,
\end{equation}
lead to fragment densities $n_\alpha(\br)=\langle \psi_\alpha|\hat{n}_\alpha(\br)|\psi_\alpha\rangle$ with the additive property:
\begin{equation}
\label{eq:den-constr}
\sum_\alpha n_\alpha(\br)=n(\br)~~,
\end{equation}
In Equation \ref{eqn:Fragment-Schroedinger}, we have defined the fragment hamiltonian with ground state $|\psi_\alpha\rangle$ as $\hat{H}_\alpha \equiv \hat{T}_\alpha+\hatV_{\sss{ee},\alpha}+\intdhrs v_{\alpha}(\br)\hat{n}_\alpha(\br)$.
Because the $n_\alpha(\br)$ are true ground-state densities for $N_\alpha$ electrons in $v_\alpha(\br)+v_p(\br)$, virial relations analogous to Equations \ref{eqn:k} and \ref{eqn:virial_KS} hold for the fragments:
\begin{equation}\label{eqn:virial_frag}
2K_\alpha+V_{\sss{ee},\alpha}=\intdhrs n_\alpha(\br)\br\cdot\nabla [v_\alpha(\br)+v\p(\br)]~~,
\end{equation}
where $K_\alpha=K[n_\alpha]=\langle\psi_\alpha[n_\alpha]|\hatT_\alpha|\psi_\alpha[n_\alpha]\rangle$, $V_{\sss{ee},\alpha}=V_{\sss{ee}}[n_\alpha]=\langle\psi_\alpha[n_\alpha]|\hatV_{\sss{ee},\alpha}|\psi_\alpha[n_\alpha]\rangle$, and
\begin{equation}\label{eqn:virial_frag_KS}
2T_{\sss{s},\alpha}=\intdhrs n_\alpha(\br)\br\cdot\nabla v_{\sss{s},\alpha}[n_\alpha](\br)~~.
\end{equation}
Note that $v_{\sss{s},\alpha}(\br)=v_\alpha(\br)+v_{\sss{H},\alpha}(\br)+v_{\sss{XC},\alpha}(\br)+v\p(\br)$. Next, subtract Equation \ref{eqn:virial_frag_KS} from Equation \ref{eqn:virial_frag} to get
\begin{equation}\label{eqn:frag_Tc}
T_{\sss{c},\alpha}=-E_{\sss{XC},\alpha}-\intdhrs n_\alpha(\br)\br\cdot\nabla v_{\sss{XC},\alpha}(\br)~~,
\end{equation}
where $T_{\sss{c},\alpha} = K_\alpha - T\s{}_{,\alpha}$ is the correlation kinetic energy of fragment $\alpha$.\cite{LP85} 

Summing up Equation \ref{eqn:virial_frag} over all fragments, we obtain
\begin{eqnarray}\label{eqn:kf}
2K\f&[\{n_\alpha\}]&+V_{\sss{ee,f}}[\{n_\alpha\}]=\nonumber\\
& &\sum_{\alpha}\intdhrs n_\alpha(\br)\br\cdot\nabla [v_\alpha(\br)+v\p(\br)]~~,
\end{eqnarray}
%
%
where $K\f[\{n_\alpha\}]\equiv\sum_{\alpha}K_\alpha$ and $V\ee{}_{,}{}\f[\{n_\alpha\}]=\sum_{\alpha}V\ee{}_{,\alpha}$.
Finally, combining Equation \ref{eqn:k} with Equation \ref{eqn:kf} and rearranging terms:
\begin{eqnarray}\label{eqn:virial_pdft}
K\nad[\{n_\alpha\}] &=& V\ext\nad[\{n_\alpha\}]+\intdhrs \sum_{\alpha}n_\alpha(\br)\br\cdot\nabla v_{{\sss ext},\alpha}\nad(\br)\nonumber\\
& &-E\p[n]-\intdhrs n(\br)\br\cdot\nabla v\p(\br)~~,
\end{eqnarray}
%
%
where $E_p[n]$ is the partition energy \cite{EBCW10}. Its functional derivative, at the minimum, is $v_p(\br)$. In Equation \ref{eqn:virial_pdft}, $V\ext\nad[\{n_{\alpha}\}]=\intdhrs\left\{n(\br)v(\br)-\sum_{\alpha}n_{\alpha}(\br)v_\alpha(\br)\right\}$ is the non-additive external energy, and $v_{{\sss ext},\alpha}\nad(\br)\equiv {\delta V\ext\nad}/{\delta n_\alpha(\br)}=v(\br)-v_\alpha(\br)$.

Equation \ref{eqn:virial_pdft} provides a way to calculate the non-additive KE in terms of quantities that can all be obtained through embedding (P-DFT) calculations.

Alternatively, subtracting Equation \ref{eqn:virial_frag_KS} from Equation \ref{eqn:virial_KS}, we find:
\begin{equation}\label{eqn:virial_pdft_alt}
T\nad_{\sss s}[\{n_\alpha\}]=\half\intdhrs\{\sum_{\alpha}n_\alpha(\br)\br\cdot\nabla [v\s(\br)-v_{\sss{s},\alpha}(\br)]\}~~,
\end{equation}
providing, together with Equation \ref{eqn:virial_pdft}, a route to the calculation of the non-additive correlation kinetic energy, as 
\begin{equation}
\label{eqn:Knad}
T_{\sss c}\nad [\{n_\alpha\}] = K\nad[\{n_\alpha\}]-T_{\sss s}^{\rm nad} [\{n_\alpha\}]~~.
\end{equation}

\section{Numerical Verification and Discussion}
\label{sec:3}
In Tables \ref{tab:virial1} and \ref{tab:virial2}, we provide numerical verification of Equations \ref{eqn:virial_pdft} and \ref{eqn:virial_pdft_alt} on diatomic molecules (i.e. each molecule is partitioned into its two constituent atoms). All calculations are performed on a real-space code that solves the KS equations in 
prolate spheroidal coordinates \cite{NW14}. P-DFT calculations were done with an algorithm that is numerically ``exact” for a given approximation to the XC functional \cite{NJW17}.

Table \ref{tab:virial1} shows a very close agreement between the non-additive kinetic energy calculated through Equation \ref{eqn:Knad_def}, denoted as $K\nad_{\sss I}$, and calculated through the virial relation of Equation \ref{eqn:virial_pdft}, denoted as $K\nad_{\sss II}$.
The main source of error comes from the calculation of the gradient of the potentials on the right-hand-side of Equation \ref{eqn:virial_pdft}, as the densities have cusps and the potentials singularities at the nuclei. Similar agreement can be seen in Table \ref{tab:virial2} that compares $T_{s,{\rm I}}^{\rm nad}$ (Equation \ref{e:Ts_I}) and $T_{s,{\rm II}}^{\rm nad}$ (Equation \ref{eqn:virial_pdft_alt}).


\begin{table}
\caption{{Numerical verification of Equation \ref{eqn:virial_pdft}. $K_{\rm I}^{\rm nad}$ is calculated through Equations \ref{eqn:Knad_def}, \ref{eqn:Knad}, and \ref{eqn:virial_pdft_alt}. $K_{\rm II}^{\rm nad}$ is calculated through Equation \ref{eqn:virial_pdft}. The H$_2^+$ result in the top line is from exact one-electron calculations for which $K_{\rm I}^{\rm nad}$ is calculated directly from wavefunctions.}}
\label{tab:virial1}
\begin{center}
\begin{tabular}{c c c}
\hline
	{System}     & {$K\nad_{\sss I}\times10^2$}  & {$K\nad_{\sss I}/K\nad_{\sss II}$} \\
\hline
	H$_2^+$ (exact)& -8.522 & 0.99914 \\
	H$_2^+$ & -8.259  & 0.99994 \\
	H$_2$  & -12.571  & 0.99993 \\
	Li$_2$ & 1.716    & 1.01237 \\
	He$_2$ & 0.1107    & 1.00025 \\
	Ne$_2$ & 0.2999   & 1.00366 \\
	Ar$_2$ & 0.4424    & 1.00417 \\
\hline
\end{tabular}	
\end{center}
\end{table}

The results in Tables \ref{tab:virial1} and \ref{tab:virial2} are a numerical verification of Equations \ref{eqn:virial_pdft} and \ref{eqn:virial_pdft_alt}. The virial relation is satisfied for each fragment \textit{and} for the full molecule. The latter occurs because the algorithm of ref.\cite{NJW17} guarantees that the sum of fragment densities reproduces the full-molecular KS density.


\begin{table}
\caption{Numerical verification of Equation \ref{eqn:virial_pdft_alt}. $T\nad_{\sss s,I}$ is calculated through Equation \ref{e:Ts_I}, and $T\nad_{\sss s,II}$ is calculated through Equation \ref{eqn:virial_pdft_alt}.}
\label{tab:virial2}
\begin{center}
\begin{tabular}{c c c}
\hline
{System}     & {$T_{s,{\rm I}}^{\rm nad}\times10^2$} & {$T_{s,{\rm I}}^{\rm nad}/T_{s,{\rm II}}^{\rm nad}$} \\
\hline
H$_2^+$ & -8.181 & 0.99997 \\
H$_2$  & -15.207  & 0.99997 \\
Li$_2$ & 0.4917    & 1.0035 \\
He$_2$ & 0.0993         & 1.00014 \\
Ne$_2$ & 0.2750    & 1.00239 \\
Ar$_2$ & 0.4050     & 1.00239 \\
\hline
\end{tabular}
\end{center}	
\end{table}

%
%
Table \ref{tab:virial_approx} provides the ratio $T_{{\sss{s}},{\rm I}}^{\rm nad}/T_{{\sss{s}},{\rm II}}^{\rm nad}$ for He$_2$ when instead of using exact numerical inversions, as before, one uses an approximate density-functional for $T_{\sss{s}}^{\rm nad}$, as is typically done in subsystem-DFT calculations \cite{KSGP15}. $T_{{\sss{s}},{\rm I}}^{\rm nad}$ is constructed from approximate $T_{\sss{s}}[n]$ functionals on the right-hand side of Equation \ref{e:Ts_I}. In Equation \ref{eqn:virial_pdft_alt}, $T\nad_{\sss s,II}$ is calculated with the same $T\s[n]$ approximation and the expression $v\s(\br)-v_{\sss{s},\alpha}(\br)=\delta T\s^{\rm nad}[n]/\delta n_{\alpha}(\br)$ \cite{NJW17}.
In contrast to the results of the exact inversion algorithm, Equations \ref{eqn:virial_pdft} and \ref{eqn:virial_pdft_alt} are not trivially satisfied in the case of approximate density functionals. The full-molecular density $n(\br)$, resulting from the sum of fragment densities in Equation \ref{eq:den-constr}, is now a self-consistent result and does not reproduce the full-molecular KS density.

\begin{table}
\caption{Comparison in the NAKE of He$_2$ when approximated functionals are used. $T\nad_{\sss s,I}$ is calculated directly from the approximated functionals. $T\nad_{\sss s,II}$ is calculated using Equation \ref{eqn:virial_pdft_alt}, where the approximated NAKE functionals are used in calculating the partition potential.}
\label{tab:virial_approx}
\begin{center}
\begin{tabular}{c c c c}	
\hline
Functional     & $T_{s,{\rm I}}^{\rm nad}\times 10^3$ & $T_{s,{\rm II}}^{\rm nad}\times 10^3$ & {$T_{s,{\rm I}}^{\rm nad}$/$T_{s,{\rm II}}^{\rm nad}$} \\
\hline
TF\cite{Tho26,Fer27}   & 1.198  & 1.402  & 0.85419 \\
vW\cite{Wei35}   & -37.823 & -37.824 & 0.99999 \\
GEA2\cite{KP56,Kir57a}   & -1.561 & -1.146 & 1.36154 \\
TW02\cite{TW02a} & 1.136  & 1.444   & 0.78654 \\
LC94\cite{LC94}  & 0.565  & 0.812  & 0.69630 \\
R-PBE\cite{JNW18} & 0.995  & 1.196  & 0.83215 \\
\hline
\end{tabular}	
\end{center}
\end{table}

For most approximate $T_{{\sss{s}}}^{\rm nad}$ functionals, the virial relation Equation \ref{eqn:virial_pdft_alt} is not well preserved.
Interestingly, the von Weis\"{a}cker (vW) functional yields an extremely accurate virial relation for He$_2$ even though the vW functional is only exact for the fragments.
This indicates that the left-hand side and the right-hand side of Equation \ref{eqn:virial_KS} are nearly equal for this approximate functional, implying that
$T\s\VW[n^{\sss{\VWa,P-DFT}}] = T\s[\tilde{n}]$,
where $n^{\sss{\VWa,P-DFT}}$ is the He$_2$ density from a P-DFT calculation that uses the vW functional, and $\tilde{n}$ is the exact density corresponding to the KS potential $v\s[n^{\sss{\VWa,P-DFT}}](\br)$, where $v\s[n^{\sss{\VWa,P-DFT}}](\br)$ is calculated by plugging the density $n^{\sss{\VWa,P-DFT}}$ into the Hartree, XC and external potential functionals. The entire error here is fragment-density-driven \cite{WNJKSB17}, and it is clearly very small. However, our previous study \cite{JNW18} showed that vW NAKE performed poorly for systems of rare gas dimers, indicating that the performance of NAKE functionals should not be judged based on Equation \ref{eqn:virial_pdft_alt} \textit{alone}.

\section{Concluding Remark}

We derived two virial relations for fragments in molecules. The first, Equation \ref{eqn:virial_pdft}, refers to the real, physical system of interacting electrons and the second, Equation \ref{eqn:virial_pdft_alt}, to the auxiliary system of non-interacting electrons.
Numerical calculations verify both relations when the exact $T_{\sss s}^{\rm nad}$ is employed and self-consistency for the fragments is reached.
These relations can be used as tools to test
approximations for $T_{\sss s}^{\rm nad}[\{n_\alpha\}]$ as a functional of the fragment densities.

\section*{Acknowledgements}
This paper is based upon work supported by the National Science Foundation under Grant No. CHE-1900301

\section*{Conflict of interest}
You may be asked to provide a conflict of interest statement during the submission process. Please check the journal's author guidelines for details on what to include in this section. Please ensure you liaise with all co-authors to confirm agreement with the final statement.


\bibliography{JNW19%
}

\begin{thebibliography}{34}%
\makeatletter
\providecommand \@ifxundefined [1]{%
 \@ifx{#1\undefined}
}%
\providecommand \@ifnum [1]{%
 \ifnum #1\expandafter \@firstoftwo
 \else \expandafter \@secondoftwo
 \fi
}%
\providecommand \@ifx [1]{%
 \ifx #1\expandafter \@firstoftwo
 \else \expandafter \@secondoftwo
 \fi
}%
\providecommand \natexlab [1]{#1}%
\providecommand \enquote  [1]{``#1''}%
\providecommand \bibnamefont  [1]{#1}%
\providecommand \bibfnamefont [1]{#1}%
\providecommand \citenamefont [1]{#1}%
\providecommand \href@noop [0]{\@secondoftwo}%
\providecommand \href [0]{\begingroup \@sanitize@url \@href}%
\providecommand \@href[1]{\@@startlink{#1}\@@href}%
\providecommand \@@href[1]{\endgroup#1\@@endlink}%
\providecommand \@sanitize@url [0]{\catcode `\\12\catcode `\$12\catcode
  `\&12\catcode `\#12\catcode `\^12\catcode `\_12\catcode `\%12\relax}%
\providecommand \@@startlink[1]{}%
\providecommand \@@endlink[0]{}%
\providecommand \url  [0]{\begingroup\@sanitize@url \@url }%
\providecommand \@url [1]{\endgroup\@href {#1}{\urlprefix }}%
\providecommand \urlprefix  [0]{URL }%
\providecommand \Eprint [0]{\href }%
\providecommand \doibase [0]{http://dx.doi.org/}%
\providecommand \selectlanguage [0]{\@gobble}%
\providecommand \bibinfo  [0]{\@secondoftwo}%
\providecommand \bibfield  [0]{\@secondoftwo}%
\providecommand \translation [1]{[#1]}%
\providecommand \BibitemOpen [0]{}%
\providecommand \bibitemStop [0]{}%
\providecommand \bibitemNoStop [0]{.\EOS\space}%
\providecommand \EOS [0]{\spacefactor3000\relax}%
\providecommand \BibitemShut  [1]{\csname bibitem#1\endcsname}%
\let\auto@bib@innerbib\@empty
\bibitem [{\citenamefont {Sun}\ and\ \citenamefont {Chan}(2016)}]{SC16}%
  \BibitemOpen
  \bibfield  {author} {\bibinfo {author} {\bibfnamefont {Q.}~\bibnamefont
  {Sun}}\ and\ \bibinfo {author} {\bibfnamefont {G.~K.-L.}\ \bibnamefont
  {Chan}},\ }\href@noop {} {\bibfield  {journal} {\bibinfo  {journal} {Acc.
  Chem. Res.}\ }\textbf {\bibinfo {volume} {49}} (\bibinfo {year}
  {2016})}\BibitemShut {NoStop}%
\bibitem [{\citenamefont {Nafziger}\ and\ \citenamefont
  {Wasserman}(2014)}]{NW14}%
  \BibitemOpen
  \bibfield  {author} {\bibinfo {author} {\bibfnamefont {J.}~\bibnamefont
  {Nafziger}}\ and\ \bibinfo {author} {\bibfnamefont {A.}~\bibnamefont
  {Wasserman}},\ }\href {\doibase 10.1021/jp504058s} {\bibfield  {journal}
  {\bibinfo  {journal} {J. Phys. Chem. A}\ }\textbf {\bibinfo {volume} {118}},\
  \bibinfo {pages} {7623} (\bibinfo {year} {2014})}\BibitemShut {NoStop}%
\bibitem [{\citenamefont {Gomez}\ \emph {et~al.}(2019)\citenamefont {Gomez},
  \citenamefont {Oueis}, \citenamefont {Restrepo},\ and\ \citenamefont
  {Wasserman}}]{GORW19}%
  \BibitemOpen
  \bibfield  {author} {\bibinfo {author} {\bibfnamefont {S.}~\bibnamefont
  {Gomez}}, \bibinfo {author} {\bibfnamefont {Y.}~\bibnamefont {Oueis}},
  \bibinfo {author} {\bibfnamefont {A.}~\bibnamefont {Restrepo}}, \ and\
  \bibinfo {author} {\bibfnamefont {A.}~\bibnamefont {Wasserman}},\ }\href@noop
  {} {\bibfield  {journal} {\bibinfo  {journal} {Int. J. Quantum Chem.}\
  }\textbf {\bibinfo {volume} {119}},\ \bibinfo {pages} {e25814} (\bibinfo
  {year} {2019})}\BibitemShut {NoStop}%
\bibitem [{\citenamefont {Huang}\ \emph {et~al.}(2014)\citenamefont {Huang},
  \citenamefont {Libisch}, \citenamefont {Peng},\ and\ \citenamefont
  {Carter}}]{HLPC14}%
  \BibitemOpen
  \bibfield  {author} {\bibinfo {author} {\bibfnamefont {C.}~\bibnamefont
  {Huang}}, \bibinfo {author} {\bibfnamefont {F.}~\bibnamefont {Libisch}},
  \bibinfo {author} {\bibfnamefont {Q.}~\bibnamefont {Peng}}, \ and\ \bibinfo
  {author} {\bibfnamefont {E.~A.}\ \bibnamefont {Carter}},\ }\href@noop {}
  {\bibfield  {journal} {\bibinfo  {journal} {J. Chem. Phys.}\ }\textbf
  {\bibinfo {volume} {140}},\ \bibinfo {pages} {124113} (\bibinfo {year}
  {2014})}\BibitemShut {NoStop}%
\bibitem [{\citenamefont {Genova}\ \emph {et~al.}(2014)\citenamefont {Genova},
  \citenamefont {Ceresoli},\ and\ \citenamefont {Pavanello}}]{GCP14}%
  \BibitemOpen
  \bibfield  {author} {\bibinfo {author} {\bibfnamefont {A.}~\bibnamefont
  {Genova}}, \bibinfo {author} {\bibfnamefont {D.}~\bibnamefont {Ceresoli}}, \
  and\ \bibinfo {author} {\bibfnamefont {M.}~\bibnamefont {Pavanello}},\ }\href
  {\doibase 10.1063/1.4897559} {\bibfield  {journal} {\bibinfo  {journal} {J.
  Chem. Phys.}\ }\textbf {\bibinfo {volume} {141}},\ \bibinfo {pages} {174101}
  (\bibinfo {year} {2014})}\BibitemShut {NoStop}%
\bibitem [{\citenamefont {Hohenberg}\ and\ \citenamefont {Kohn}(1964)}]{HK64}%
  \BibitemOpen
  \bibfield  {author} {\bibinfo {author} {\bibfnamefont {P.}~\bibnamefont
  {Hohenberg}}\ and\ \bibinfo {author} {\bibfnamefont {W.}~\bibnamefont
  {Kohn}},\ }\href {\doibase 10.1103/PhysRev.136.B864} {\bibfield  {journal}
  {\bibinfo  {journal} {Phys. Rev.}\ }\textbf {\bibinfo {volume} {136}},\
  \bibinfo {pages} {B864} (\bibinfo {year} {1964})}\BibitemShut {NoStop}%
\bibitem [{\citenamefont {Kohn}\ and\ \citenamefont {Sham}(1965)}]{KS65}%
  \BibitemOpen
  \bibfield  {author} {\bibinfo {author} {\bibfnamefont {W.}~\bibnamefont
  {Kohn}}\ and\ \bibinfo {author} {\bibfnamefont {L.~J.}\ \bibnamefont
  {Sham}},\ }\href {\doibase 10.1103/PhysRev.140.A1133} {\bibfield  {journal}
  {\bibinfo  {journal} {Phys. Rev.}\ }\textbf {\bibinfo {volume} {140}},\
  \bibinfo {pages} {A1133} (\bibinfo {year} {1965})}\BibitemShut {NoStop}%
\bibitem [{\citenamefont {Weso{\l}owski}\ and\ \citenamefont
  {Warshel}(1993)}]{WW93}%
  \BibitemOpen
  \bibfield  {author} {\bibinfo {author} {\bibfnamefont {T.~A.}\ \bibnamefont
  {Weso{\l}owski}}\ and\ \bibinfo {author} {\bibfnamefont {A.}~\bibnamefont
  {Warshel}},\ }\href@noop {} {\bibfield  {journal} {\bibinfo  {journal} {J.
  Phys. Chem.}\ }\textbf {\bibinfo {volume} {97}},\ \bibinfo {pages} {8050}
  (\bibinfo {year} {1993})}\BibitemShut {NoStop}%
\bibitem [{\citenamefont {Krishtal}\ \emph {et~al.}(2015)\citenamefont
  {Krishtal}, \citenamefont {Sinha}, \citenamefont {Genova},\ and\
  \citenamefont {Pavanello}}]{KSGP15}%
  \BibitemOpen
  \bibfield  {author} {\bibinfo {author} {\bibfnamefont {A.}~\bibnamefont
  {Krishtal}}, \bibinfo {author} {\bibfnamefont {D.}~\bibnamefont {Sinha}},
  \bibinfo {author} {\bibfnamefont {A.}~\bibnamefont {Genova}}, \ and\ \bibinfo
  {author} {\bibfnamefont {M.}~\bibnamefont {Pavanello}},\ }\href@noop {}
  {\bibfield  {journal} {\bibinfo  {journal} {J. Phys. Condens. Matter}\
  }\textbf {\bibinfo {volume} {27}},\ \bibinfo {pages} {183202} (\bibinfo
  {year} {2015})}\BibitemShut {NoStop}%
\bibitem [{\citenamefont {Cohen}\ and\ \citenamefont {Wasserman}(2007)}]{CW07}%
  \BibitemOpen
  \bibfield  {author} {\bibinfo {author} {\bibfnamefont {M.~H.}\ \bibnamefont
  {Cohen}}\ and\ \bibinfo {author} {\bibfnamefont {A.}~\bibnamefont
  {Wasserman}},\ }\href {\doibase 10.1021/jp066449h} {\bibfield  {journal}
  {\bibinfo  {journal} {J. Phys. Chem. A}\ }\textbf {\bibinfo {volume} {111}},\
  \bibinfo {pages} {2229} (\bibinfo {year} {2007})}\BibitemShut {NoStop}%
\bibitem [{\citenamefont {Elliott}\ \emph {et~al.}(2010)\citenamefont
  {Elliott}, \citenamefont {Burke}, \citenamefont {Cohen},\ and\ \citenamefont
  {Wasserman}}]{EBCW10}%
  \BibitemOpen
  \bibfield  {author} {\bibinfo {author} {\bibfnamefont {P.}~\bibnamefont
  {Elliott}}, \bibinfo {author} {\bibfnamefont {K.}~\bibnamefont {Burke}},
  \bibinfo {author} {\bibfnamefont {M.~H.}\ \bibnamefont {Cohen}}, \ and\
  \bibinfo {author} {\bibfnamefont {A.}~\bibnamefont {Wasserman}},\ }\href
  {\doibase 10.1103/PhysRevA.82.024501} {\bibfield  {journal} {\bibinfo
  {journal} {Phys. Rev. A}\ }\textbf {\bibinfo {volume} {82}},\ \bibinfo
  {pages} {024501} (\bibinfo {year} {2010})}\BibitemShut {NoStop}%
\bibitem [{\citenamefont {Manby}\ \emph {et~al.}(2012)\citenamefont {Manby},
  \citenamefont {Stella}, \citenamefont {Goodpaster},\ and\ \citenamefont
  {Miller~III}}]{MSGMI12}%
  \BibitemOpen
  \bibfield  {author} {\bibinfo {author} {\bibfnamefont {F.~R.}\ \bibnamefont
  {Manby}}, \bibinfo {author} {\bibfnamefont {M.}~\bibnamefont {Stella}},
  \bibinfo {author} {\bibfnamefont {J.~D.}\ \bibnamefont {Goodpaster}}, \ and\
  \bibinfo {author} {\bibfnamefont {T.~F.}\ \bibnamefont {Miller~III}},\
  }\href@noop {} {\bibfield  {journal} {\bibinfo  {journal} {J. Chem. Theory
  Comput.}\ }\textbf {\bibinfo {volume} {8}},\ \bibinfo {pages} {2564}
  (\bibinfo {year} {2012})}\BibitemShut {NoStop}%
\bibitem [{\citenamefont {Wesolowski}\ and\ \citenamefont {Wang}(2013)}]{WW13}%
  \BibitemOpen
  \bibfield  {author} {\bibinfo {author} {\bibfnamefont {T.~A.}\ \bibnamefont
  {Wesolowski}}\ and\ \bibinfo {author} {\bibfnamefont {Y.~A.}\ \bibnamefont
  {Wang}},\ }\href@noop {} {\emph {\bibinfo {title} {Recent Progress in
  Orbital-free Density Functional Theory}}}\ (\bibinfo  {publisher} {WORLD
  SCIENTIFIC},\ \bibinfo {year} {2013})\BibitemShut {NoStop}%
\bibitem [{\citenamefont {K\"{u}mmel}\ and\ \citenamefont
  {Kronik}(2008)}]{KK08}%
  \BibitemOpen
  \bibfield  {author} {\bibinfo {author} {\bibfnamefont {S.}~\bibnamefont
  {K\"{u}mmel}}\ and\ \bibinfo {author} {\bibfnamefont {L.}~\bibnamefont
  {Kronik}},\ }\href@noop {} {\bibfield  {journal} {\bibinfo  {journal} {Rev.
  Mod. Phys.}\ }\textbf {\bibinfo {volume} {80}},\ \bibinfo {pages} {3}
  (\bibinfo {year} {2008})}\BibitemShut {NoStop}%
\bibitem [{\citenamefont {Genova}\ \emph {et~al.}(2016)\citenamefont {Genova},
  \citenamefont {Ceresoli},\ and\ \citenamefont {Pavanello}}]{GCP16}%
  \BibitemOpen
  \bibfield  {author} {\bibinfo {author} {\bibfnamefont {A.}~\bibnamefont
  {Genova}}, \bibinfo {author} {\bibfnamefont {D.}~\bibnamefont {Ceresoli}}, \
  and\ \bibinfo {author} {\bibfnamefont {M.}~\bibnamefont {Pavanello}},\
  }\href@noop {} {\bibfield  {journal} {\bibinfo  {journal} {J. Chem. Phys.}\
  }\textbf {\bibinfo {volume} {144}},\ \bibinfo {pages} {234105} (\bibinfo
  {year} {2016})}\BibitemShut {NoStop}%
\bibitem [{\citenamefont {Lastra}\ \emph {et~al.}(2008)\citenamefont {Lastra},
  \citenamefont {Kaminski},\ and\ \citenamefont {Weso{\l}owski}}]{LKW08}%
  \BibitemOpen
  \bibfield  {author} {\bibinfo {author} {\bibfnamefont {J.~M.~G.}\
  \bibnamefont {Lastra}}, \bibinfo {author} {\bibfnamefont {J.~W.}\
  \bibnamefont {Kaminski}}, \ and\ \bibinfo {author} {\bibfnamefont {T.~A.}\
  \bibnamefont {Weso{\l}owski}},\ }\href@noop {} {\bibfield  {journal}
  {\bibinfo  {journal} {J. Chem. Phys.}\ }\textbf {\bibinfo {volume} {129}}
  (\bibinfo {year} {2008})}\BibitemShut {NoStop}%
\bibitem [{\citenamefont {Levy}\ and\ \citenamefont {Perdew}(1985)}]{LP85}%
  \BibitemOpen
  \bibfield  {author} {\bibinfo {author} {\bibfnamefont {M.}~\bibnamefont
  {Levy}}\ and\ \bibinfo {author} {\bibfnamefont {J.~P.}\ \bibnamefont
  {Perdew}},\ }\href {\doibase 10.1103/PhysRevA.32.2010} {\bibfield  {journal}
  {\bibinfo  {journal} {Phys. Rev. A}\ }\textbf {\bibinfo {volume} {32}},\
  \bibinfo {pages} {2010} (\bibinfo {year} {1985})}\BibitemShut {NoStop}%
\bibitem [{\citenamefont {Ghosh}\ and\ \citenamefont {Singh}(1989)}]{GS89}%
  \BibitemOpen
  \bibfield  {author} {\bibinfo {author} {\bibfnamefont {S.}~\bibnamefont
  {Ghosh}}\ and\ \bibinfo {author} {\bibfnamefont {V.~A.}\ \bibnamefont
  {Singh}},\ }\href@noop {} {\bibfield  {journal} {\bibinfo  {journal} {J.
  Phys. Condens. Matter}\ }\textbf {\bibinfo {volume} {1}},\ \bibinfo {pages}
  {1971} (\bibinfo {year} {1989})}\BibitemShut {NoStop}%
\bibitem [{\citenamefont {Rodr{\'{i}}guez}\ \emph {et~al.}(2009)\citenamefont
  {Rodr{\'{i}}guez}, \citenamefont {Ayers}, \citenamefont {G{\"{o}}tz},\ and\
  \citenamefont {Castillo-Alvarado}}]{RAGC09}%
  \BibitemOpen
  \bibfield  {author} {\bibinfo {author} {\bibfnamefont {J.~I.}\ \bibnamefont
  {Rodr{\'{i}}guez}}, \bibinfo {author} {\bibfnamefont {P.~W.}\ \bibnamefont
  {Ayers}}, \bibinfo {author} {\bibfnamefont {A.~W.}\ \bibnamefont
  {G{\"{o}}tz}}, \ and\ \bibinfo {author} {\bibfnamefont {F.~L.}\ \bibnamefont
  {Castillo-Alvarado}},\ }\href {\doibase 10.1063/1.3160670} {\bibfield
  {journal} {\bibinfo  {journal} {J. Chem. Phys.}\ }\textbf {\bibinfo {volume}
  {131}},\ \bibinfo {pages} {021101} (\bibinfo {year} {2009})}\BibitemShut
  {NoStop}%
\bibitem [{\citenamefont {Srebrenik}\ and\ \citenamefont {Bader}(1975)}]{SB75}%
  \BibitemOpen
  \bibfield  {author} {\bibinfo {author} {\bibfnamefont {S.}~\bibnamefont
  {Srebrenik}}\ and\ \bibinfo {author} {\bibfnamefont {R.~F.~W.}\ \bibnamefont
  {Bader}},\ }\href@noop {} {\bibfield  {journal} {\bibinfo  {journal} {J.
  Chem. Phys.}\ }\textbf {\bibinfo {volume} {63}},\ \bibinfo {pages} {3945}
  (\bibinfo {year} {1975})}\BibitemShut {NoStop}%
\bibitem [{\citenamefont {Bader}(1986)}]{B86}%
  \BibitemOpen
  \bibfield  {author} {\bibinfo {author} {\bibfnamefont {R.~F.~W.}\
  \bibnamefont {Bader}},\ }\href@noop {} {\bibfield  {journal} {\bibinfo
  {journal} {Journal of Chemical Physics}\ }\textbf {\bibinfo {volume} {85}},\
  \bibinfo {pages} {3133} (\bibinfo {year} {1986})}\BibitemShut {NoStop}%
\bibitem [{\citenamefont {Li}\ and\ \citenamefont {Parr}(1986)}]{LP86}%
  \BibitemOpen
  \bibfield  {author} {\bibinfo {author} {\bibfnamefont {L.~M.}\ \bibnamefont
  {Li}}\ and\ \bibinfo {author} {\bibfnamefont {R.~G.}\ \bibnamefont {Parr}},\
  }\href@noop {} {\bibfield  {journal} {\bibinfo  {journal} {J. Chem. Phys.}\
  }\textbf {\bibinfo {volume} {84}},\ \bibinfo {pages} {1704} (\bibinfo {year}
  {1986})}\BibitemShut {NoStop}%
\bibitem [{\citenamefont {Harris}\ and\ \citenamefont {Heller}(1975)}]{HH75}%
  \BibitemOpen
  \bibfield  {author} {\bibinfo {author} {\bibfnamefont {R.~A.}\ \bibnamefont
  {Harris}}\ and\ \bibinfo {author} {\bibfnamefont {D.~F.}\ \bibnamefont
  {Heller}},\ }\href@noop {} {\bibfield  {journal} {\bibinfo  {journal} {J.
  Chem. Phys.}\ }\textbf {\bibinfo {volume} {62}},\ \bibinfo {pages} {3601}
  (\bibinfo {year} {1975})}\BibitemShut {NoStop}%
\bibitem [{\citenamefont {Gordon}\ and\ \citenamefont {Kim}(1972)}]{GK72}%
  \BibitemOpen
  \bibfield  {author} {\bibinfo {author} {\bibfnamefont {R.~G.}\ \bibnamefont
  {Gordon}}\ and\ \bibinfo {author} {\bibfnamefont {Y.~S.}\ \bibnamefont
  {Kim}},\ }\href@noop {} {\bibfield  {journal} {\bibinfo  {journal} {J. Chem.
  Phys.}\ }\textbf {\bibinfo {volume} {56}},\ \bibinfo {pages} {3122} (\bibinfo
  {year} {1972})}\BibitemShut {NoStop}%
\bibitem [{\citenamefont {Nafziger}\ \emph {et~al.}(2017)\citenamefont
  {Nafziger}, \citenamefont {Jiang},\ and\ \citenamefont {Wasserman}}]{NJW17}%
  \BibitemOpen
  \bibfield  {author} {\bibinfo {author} {\bibfnamefont {J.}~\bibnamefont
  {Nafziger}}, \bibinfo {author} {\bibfnamefont {K.}~\bibnamefont {Jiang}}, \
  and\ \bibinfo {author} {\bibfnamefont {A.}~\bibnamefont {Wasserman}},\
  }\href@noop {} {\bibfield  {journal} {\bibinfo  {journal} {J. Chem. Theory
  Comput.}\ }\textbf {\bibinfo {volume} {13}},\ \bibinfo {pages} {577}
  (\bibinfo {year} {2017})}\BibitemShut {NoStop}%
\bibitem [{\citenamefont {Thomas}(1926)}]{Tho26}%
  \BibitemOpen
  \bibfield  {author} {\bibinfo {author} {\bibfnamefont {L.~H.}\ \bibnamefont
  {Thomas}},\ }\href@noop {} {\bibfield  {journal} {\bibinfo  {journal} {Math.
  Proc. Cambridge Philos. Soc.}\ }\textbf {\bibinfo {volume} {23}},\ \bibinfo
  {pages} {542} (\bibinfo {year} {1926})}\BibitemShut {NoStop}%
\bibitem [{\citenamefont {Fermi}(1927)}]{Fer27}%
  \BibitemOpen
  \bibfield  {author} {\bibinfo {author} {\bibfnamefont {E.}~\bibnamefont
  {Fermi}},\ }\href@noop {} {\bibfield  {journal} {\bibinfo  {journal}
  {Rendiconti. Accademia Nazionale dei Lincei}\ }\textbf {\bibinfo {volume}
  {6}},\ \bibinfo {pages} {32} (\bibinfo {year} {1927})}\BibitemShut {NoStop}%
\bibitem [{\citenamefont {v~Weizs{\"a}cker}(1935)}]{Wei35}%
  \BibitemOpen
  \bibfield  {author} {\bibinfo {author} {\bibfnamefont {C.~F.}\ \bibnamefont
  {v~Weizs{\"a}cker}},\ }\href@noop {} {\bibfield  {journal} {\bibinfo
  {journal} {Zeitschrift F{\"u}r Physik a Hadrons and Nuclei}\ }\textbf
  {\bibinfo {volume} {96}},\ \bibinfo {pages} {431} (\bibinfo {year}
  {1935})}\BibitemShut {NoStop}%
\bibitem [{\citenamefont {Kompaneets}\ and\ \citenamefont
  {Pavlovsky}(1956)}]{KP56}%
  \BibitemOpen
  \bibfield  {author} {\bibinfo {author} {\bibfnamefont {A.~S.}\ \bibnamefont
  {Kompaneets}}\ and\ \bibinfo {author} {\bibfnamefont {E.~S.}\ \bibnamefont
  {Pavlovsky}},\ }\href@noop {} {\bibfield  {journal} {\bibinfo  {journal} {J.
  Exp. Theor. Phys.}\ }\textbf {\bibinfo {volume} {31}},\ \bibinfo {pages}
  {427} (\bibinfo {year} {1956})}\BibitemShut {NoStop}%
\bibitem [{\citenamefont {Kirzhnits}(1957)}]{Kir57a}%
  \BibitemOpen
  \bibfield  {author} {\bibinfo {author} {\bibfnamefont {D.~A.}\ \bibnamefont
  {Kirzhnits}},\ }\href@noop {} {\bibfield  {journal} {\bibinfo  {journal}
  {Journal of Experimental and Theoretical Physics}\ }\textbf {\bibinfo
  {volume} {5}},\ \bibinfo {pages} {64} (\bibinfo {year} {1957})}\BibitemShut
  {NoStop}%
\bibitem [{\citenamefont {Tran}\ and\ \citenamefont
  {Weso{\l}owski}(2002)}]{TW02a}%
  \BibitemOpen
  \bibfield  {author} {\bibinfo {author} {\bibfnamefont {F.}~\bibnamefont
  {Tran}}\ and\ \bibinfo {author} {\bibfnamefont {T.~A.}\ \bibnamefont
  {Weso{\l}owski}},\ }\href {\doibase 10.1002/qua.10306} {\bibfield  {journal}
  {\bibinfo  {journal} {Int. J. Quantum Chem.}\ }\textbf {\bibinfo {volume}
  {89}},\ \bibinfo {pages} {441} (\bibinfo {year} {2002})}\BibitemShut
  {NoStop}%
\bibitem [{\citenamefont {Lembarki}\ and\ \citenamefont
  {Chermette}(1994)}]{LC94}%
  \BibitemOpen
  \bibfield  {author} {\bibinfo {author} {\bibfnamefont {A.}~\bibnamefont
  {Lembarki}}\ and\ \bibinfo {author} {\bibfnamefont {H.}~\bibnamefont
  {Chermette}},\ }\href {\doibase 10.1103/PhysRevA.50.5328} {\bibfield
  {journal} {\bibinfo  {journal} {Phys. Rev. A}\ }\textbf {\bibinfo {volume}
  {50}},\ \bibinfo {pages} {5328} (\bibinfo {year} {1994})}\BibitemShut
  {NoStop}%
\bibitem [{\citenamefont {Jiang}\ \emph {et~al.}(2018)\citenamefont {Jiang},
  \citenamefont {Nafziger},\ and\ \citenamefont {Wasserman}}]{JNW18}%
  \BibitemOpen
  \bibfield  {author} {\bibinfo {author} {\bibfnamefont {K.}~\bibnamefont
  {Jiang}}, \bibinfo {author} {\bibfnamefont {J.}~\bibnamefont {Nafziger}}, \
  and\ \bibinfo {author} {\bibfnamefont {A.}~\bibnamefont {Wasserman}},\
  }\href@noop {} {\bibfield  {journal} {\bibinfo  {journal} {J. Chem. Phys.}\
  }\textbf {\bibinfo {volume} {148}},\ \bibinfo {pages} {104113} (\bibinfo
  {year} {2018})}\BibitemShut {NoStop}%
\bibitem [{\citenamefont {Wasserman}\ \emph {et~al.}(2017)\citenamefont
  {Wasserman}, \citenamefont {Nafziger}, \citenamefont {Jiang}, \citenamefont
  {Kim}, \citenamefont {Sim},\ and\ \citenamefont {Burke}}]{WNJKSB17}%
  \BibitemOpen
  \bibfield  {author} {\bibinfo {author} {\bibfnamefont {A.}~\bibnamefont
  {Wasserman}}, \bibinfo {author} {\bibfnamefont {J.}~\bibnamefont {Nafziger}},
  \bibinfo {author} {\bibfnamefont {K.~L.}\ \bibnamefont {Jiang}}, \bibinfo
  {author} {\bibfnamefont {M.~C.}\ \bibnamefont {Kim}}, \bibinfo {author}
  {\bibfnamefont {E.}~\bibnamefont {Sim}}, \ and\ \bibinfo {author}
  {\bibfnamefont {K.}~\bibnamefont {Burke}},\ }\href@noop {} {\bibfield
  {journal} {\bibinfo  {journal} {Annu. Rev. Phys. Chem.}\ }\textbf {\bibinfo
  {volume} {68}},\ \bibinfo {pages} {555} (\bibinfo {year} {2017})}\BibitemShut
  {NoStop}%
\end{thebibliography}%

\end{document}